\documentclass[twocolumn]{aastex63}
\usepackage{hyperref}
\usepackage[colorinlistoftodos]{todonotes}
\usepackage{amsmath}

\shorttitle{Resonant Chains of Exoplanets}

\begin{document}
\title{Resonant Chains of Exoplanets: Libration Centers for Three-body Angles}

\correspondingauthor{Jared Siegel}
\email{siegeljc@uchicago.edu}

\author[0000-0002-9337-0902]{Jared C. Siegel}
\affiliation{Department of Astronomy and Astrophysics, University of Chicago \\
5640 S Ellis Ave \\
Chicago, IL 60637, USA}

\author[0000-0003-3750-0183]{Daniel Fabrycky}
\affiliation{Department of Astronomy and Astrophysics, University of Chicago \\
5640 S Ellis Ave \\
Chicago, IL 60637, USA}

\newcommand{\N}{30}
%-------------------------abstract-------------------------
\begin{abstract}

Resonant planetary systems contain at least one planet pair with orbital periods librating at a near-integer ratio (2/1, 3/2, 4/3, etc.) and are a natural outcome of standard planetary formation theories. Systems with multiple adjacent resonant pairs are known as resonant chains and can exhibit three-body resonances ---characterized by a critical three-body angle. Here we study three-body angles as a diagnostic of resonant chains through tidally-damped N-body integrations. For each combination of the 2:1, 3:2, 4:3, and 5:4 mean motion resonances (the most common resonances in the known resonant chains), we characterize the three-body angle equilibria for several mass schemes, migration timescales, and initial separations. We find that under our formulation of the three-body angle, which does not reduce coefficients, 180$^{\circ}$ is the preferred libration center, and libration centers shifted away from 180$^{\circ}$ are associated with non-adjacent resonances. We then relate these angles to observables, by applying our general results to two transiting systems: Kepler-60 and Kepler-223. For these systems, we compare N-body models of the three-body angle to the zeroth order in $e$ approximation accessible via transit phases, used in previous publications. In both cases, we find the three-body angle during the \textit{Kepler} observing window is not necessarily indicative of the long-term oscillations and stress the role of dynamical models in investigating three-body angles. We anticipate our results will provide a useful diagnostic in the analysis of resonant chains.

\end{abstract}

%-------------------------keywords-------------------------

\section{Introduction}
\label{sec:intro}
Standard theories of planet formation predict planetary migration to be pervasive. After planets are spawned in protoplanetary disks, they create gravitational wakes, which torque the planets' orbits \citep[type-I migration,][]{1997Ward}. Since the migration rate depends on the planetary mass, and moreover on the disk surface density and structure, as planets spiral in towards the star they can catch up to one another. Before overtaking each other, the planets interact strongly through resonant effects, which can transfer angular momentum among planets on a timescale even faster than the disk torques each planet. This effect gives rise to resonant trapping phenomena, in which planets migrate in lock-step, with near integer-ratio orbital periods. The eventual draining and photoevaporation of the disk can leave behind resonant planets as a fossil record of this evolution. 

Systems with multiple planet pairs each locked in a mean motion resonance (MMR) ---where the pair's orbital periods librate at a near-integer ratio--- are known as resonant chains. Simulations of planetary migration initially predicted such systems \citep{Cresswell2006}, and \cite{Mills2016} successfully reproduced the observed state of Kepler-223, a four planet resonant chain, using a gas-disk migration model. More recently, \cite{MacDonald2018} demonstrated multiple channels of migration ---long-scale, short-scale, and pure eccentricity damping--- are compatible with several observed chains.

There are now five confirmed exoplanet resonant chains, including systems composed of giants \citep[GJ 876,][]{Rivera2010}, sub-Neptunes \citep[Kepler-223,][]{Mills2016}, super-Earths with and without gas envelopes \citep[Kepler-80 and TOI-178,][]{MacDonald2016, 2021Leleu}, and terrestrials \citep[TRAPPIST-1,][]{Gillon2016,2017Luger}. In addition, there are currently three suspected resonant chains: HR 8799 \citep{Fabrycky2010}, Kepler-60 \citep{Jontof-Hutter2016, godz2016}, and K2-138 \citep{Christiansen2018, Lopez2019}; however, \cite{Jontof-Hutter2016} was recently revised by \cite{Jontof_Hutter_2021}, showing the three-body angle of Kepler-60 indeed librates, potentially promoting this system to a confirmed resonant chain. Among all these systems, period ratios between adjacent planets are predominately near first-order commensurabilities, with the only higher-order commensurabilities (5:3 and 8:5) found in the inner two pairs of TRAPPIST-1. Other systems --- e.g., HD 158259 \citep{2020Hara} and Kepler-444 \citep{2016Papaloizou} --- bear the resonant hallmarks of convergent migration, though they lack librating three-body resonances, described next. 

\subsection{Three-body Angles}
\label{sec:formulation}

In resonant systems, three-body resonances can engage and link the planets' dynamics together. Such resonances are characterized by the general three-body angle, which is found through a linear-combination of the planets' mean longitudes. These angles are easily detected in transit data, because they strongly rely on the observed transit phase (with a weak dependence on eccentricity) and are a powerful diagnostic of system architecture. 

A general three-planet resonant chain obeys $\frac{P_{2}}{P_{1}} \approx \frac{j+1}{j}$ and $\frac{P_{3}}{P_{2}} \approx \frac{k+1}{k}$, where $j,k$ are integers and $P_i$ are the periods of subsequent planets ($i=1,2,3$); we refer to such chains by $(j+1:j,k+1:k)$. For a given chain, the system can be characterized by the two critical angles:
\begin{eqnarray}
    \phi_{12} &= (j+1)\lambda_2 - j \lambda_1 - \varpi_2\\
    \phi_{23} &= (k+1)\lambda_3 - k \lambda_2 - \varpi_2 
\end{eqnarray}
where $\lambda_i$ and $\varpi_i$ are the mean longitudes and pericenter longitudes, respectively.  Subtracting $\phi_{12}$ from $\phi_{23}$ then yields the general three-body angle:
\begin{equation}
    \label{equ:tb}
    \phi = j \lambda_1 - (j+k+1)\lambda_2 +  (k+1)\lambda_3
\end{equation}
If both critical angles librate, it follows that $\phi$ will librate. This scenario, called type-I by \cite{godz2016}, is the primary focus of this paper; however, we note $\phi$ can librate even if all the two-body critical angles circulate.

Prior studies often reduce equation \ref{equ:tb} by the greatest common denominator. Unless otherwise stated, we do not reduce the angles and use the formulation of equation \ref{equ:tb}.

\subsection{Motivation and Overview}

As the number of known resonant chains grows, it has typically fallen to complex transit-timing variation (TTV) analysis \citep[Kepler-60,][]{Migaszewski2016, Jontof-Hutter2016}, involved analytics \citep[HR8799,][]{Gozdziewski2020}, or numerical modeling \citep[Kepler-223 and TRAPPIST-1,][]{Mills2016, Tamayo2017} to predict the three-body angles of observed systems.

Using a semi-analytic approach, \citet{delisle2017} developed a model of three-body angle equilibria and applied it to Kepler-223. \cite{delisle2017} found $\phi=180^{\circ}$ in cases without first-order non-adjacent MMR and identified multiple equilibria otherwise. However, this model does not treat higher-order non-adjacent MMR and requires substantial mathematical machinery, limiting its applicability to observed systems.

In a method first employed by \cite{Mills2016} and generalized by \cite{2017Luger}, the observed transit times can be mapped to mean longitudes, and by extension the three-body angle, using a zeroth order in $e$ approximation. This approach is independent of dynamical modeling and treats the three-body angle as a direct observable, however, higher-order terms may be non-negligible. 

Here we perform tidally-damped N-body integrations, in order to characterize the three-body angle equilibria for an array of resonant configurations; since the observed resonant chains predominately host planet pairs near first-order commensurabilities, we focus on combinations of the 2:1, 3:2, 4:3, and 5:4 MMRs. Motivated by the observed chains, we also focus on the super-Earth to sub-Neptune mass regime. Through these simulations and a standardized formulation of the three-body angle, we present a general rule for three-body angle equilibria. We then relate our numerical results with their observational equivalents, by conducting transit timing variation studies of Kepler-60 and Kepler-223. 

In \S \ref{sec:simulations}, we present the framework of our integrations and outline the simulations conducted in this study. Our numerical results are then presented in \S \ref{sec:sim_res}. In \S \ref{sec:application}, we apply these results to observed systems, and in \S \ref{sec:concl}, we summarize our conclusions.

\begin{deluxetable}{llccccc}[t]
\tablecaption{ Simulation Parameters \label{tab:res_grid_par}}
\tablehead{
\colhead{Name} & \colhead{Mass Scheme} & 
\colhead{$\tau_a$} &
\colhead{$K$}\\
% \colhead{$t_2$}\\
\colhead{} & 
% \colhead{} &
\colhead{($M_{\oplus}$)} &
\colhead{(days)} &
\colhead{} &
% \colhead{(days)} & 
% \colhead{}
}
\startdata
Mass Expl. & & & &\\ 
---$\mathcal{S}_1$ & 4.0, 4.0, 4.0 & $10^8$ & $10^3$\\
---$\mathcal{S}_2$ & 8.0, 8.0, 8.0 & $10^8$ & $10^3$\\
---$\mathcal{S}_3$ & 2.0, 4.0, 6.0 & $10^8$ & $10^3$\\
---$\mathcal{S}_4$ & 4.0, 6.0, 8.0 & $10^8$ & $10^3$\\
\hline
Timescale Expl. & & & &\\
---$\mathcal{S}_5$ & 4.0, 4.0, 4.0 & $10^6-10^9$ & $5-10^3$\\
\hline
\enddata
\tablecomments{ For $\mathcal{S}_1$ through $\mathcal{S}_5$, we generate models for all combinations of the 2:1, 3:2, 4:3, and 5:4 mean motion resonances. $t_1=1.46 \times 10^7$ days and $t_2=5.84\times10^7$ days for all integrations. }

\end{deluxetable}

\section{Simulations}
\label{sec:simulations}
Using tidally-damped N-body integrations, we characterize the three-body angle equilibria for an array of resonant configurations. Here we outline the methods of those integrations and summarize the simulation parameters.

\subsection{Methods}
\label{sec:integrator}
We follow the methods of \cite{MacDonald2016} and model Newtonian N-body dynamics using an 8th/9th order Prince-Dormand integration method from the GNU Scientific Library \citep{galassi2009}.  To simulate disk migration, we damp the semi-major axis (with a timescale of $\tau_a$) of the outermost planet and eccentricity (with a timescale of $\tau_e$) of every planet. Dissipation is implemented by applying an acceleration to dampen the radial and tangential velocities of individual planets with respect to the host star \citep{Thommes2008}\footnote{{ \citet{Thommes2008} mistakenly reported an extra $1/r$ factor in the $a_r$ equation.}} 
\begin{eqnarray}
a_{\phi} &=& - v_\phi / (2 \tau_a), \label{eq:ta} \\
a_r &=& -2 v_r / \tau_e, \label{eq:te}
\end{eqnarray}
where $\vec{a}$ and $\vec{v}$ are the acceleration applied to the planet and the planet's instantaneous velocity, respectively, and their components are written with subscripts, referring to a polar coordinate system where $r$ and $\phi$ are defined with respect to the central star. 
This scheme captures the planets sequentially into the resonances. Once each chain has captured into resonance and reached equilibrium eccentricity, the migration force is terminated at a time $t_1$; for every model in this study, we let $t_1=1.46\times10^7$~days. Our method brings several planets into resonance in a controlled fashion, though we are not attempting to model the likely physical situation of multiplanet migration near the inner edge of a gaseous disk, as others have done \citep{2018Brasser}; our particular choice can be related to the more general case, in which multiple planets have semi-major axis damping \citep{2010Rein}.

Following the migration phase, eccentricity damping is applied to each planet in accordance with tidal-damping, parameterized by the tidal-dissipation coefficient $Q$ \citep{Goldreich1966}, for a total time $t_2$. In general, tidal-damping suppresses a given planet's eccentricity $e$ and spreads the period ratios away from commensurability. We relate the eccentricity-damping timescale to tidal-dissipation via the formulation of  \cite{2008Jackson},
\begin{equation}
\frac{1}{\tau_e} = \frac{1}{e} \frac{de}{dt}  = -\frac{63}{4} (G M_\star)^{1/2} \frac{M_\star}{M_p} \frac{R_p^5}{a^{13/2}} \frac{1}{Q^{\prime}} + \big( \frac{1}{e} \frac{de}{dt} \big)_\star  \label{eqn:tidaldiss}
\end{equation}
where $Q^{\prime}$ is defined as,
\begin{equation}
Q^{\prime} = \frac{3Q}{2k_2},
\end{equation}
$G$ is Newton's gravitational constant, $M_\star$ and $M_p$ are the stellar and planetary masses, respectively, $R_p$ is the planetary radius, $k_2$ is the Love number quantifying the susceptibility of a planet to tidal-distortion \citep{Papaloizou2015}, and $a$ is the semi-major axis. In equation~\ref{eqn:tidaldiss}, the term $\big( \frac{1}{e} \frac{de}{dt} \big)_\star$ denotes the dissipation effect of tides raised on the star; since we are focusing on very low-mass planets, we omit that portion in our analysis. 

Unless otherwise stated, we assume $Q^{\prime}$ is constant among all planets in a system and set $Q^{\prime}$ such that the eccentricity-damping timescale of the innermost planet is ten-times longer in the tidal-damping phase than the migration phase. The eccentricity-damping of the outer planets scale according to equation \ref{eqn:tidaldiss}, assuming uniform planet radii. 

\begin{figure*}[]
\includegraphics[width=\textwidth, clip, trim=9.5cm 0.5cm 9.5cm 0.5cm]{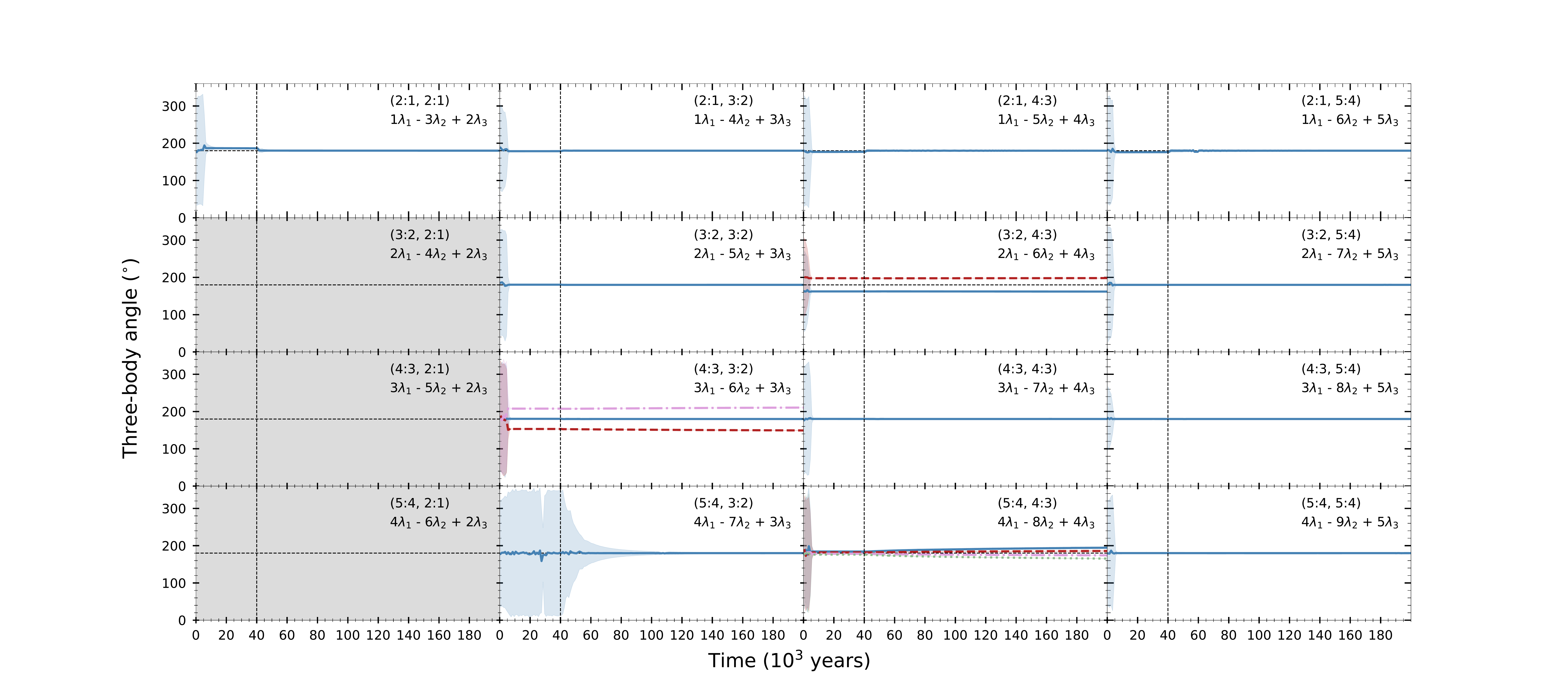}
\caption{Evolution of the three-body angles in $\mathcal{S}_1$, where $\tau_a=10^8$~days and $K=10^3$; the three-body angles are dependant on the adopted migration timescales (see \S \ref{sec:tau_dep}). For each resonant configuration, the 30 models are grouped based on their three-body angle. The mean libration center of each group is represented as a line, the mean
libration amplitude is shown as a shaded region, and the colors correspond to the different groups. The vertical dashed-black line
demarcates the end of the migration regime and the start of the tidal-dissipation forces. Configurations without any libration of the three-body angle
are shaded grey. }
\label{fig:tb}
\end{figure*}

\subsection{Simulation Parameters}
\label{sec:sim_par}

In order to characterize the three-body angle equilibria, we generate and integrate synthetic systems for each chain of the form $(j+1:j,k+1:k)$, where $j,k \in \{1,2,3,4\}$. For clarity, we group our models into two sets: mass exploration and timescale exploration. For the mass exploration suites, we are interested in mapping out the resonant equilibria of each configuration and comparing between four different planet mass schemes. With the timescale exploration suite, we investigate the sensitivity of these equilibria to a wide range of migration timescales and initial positions. Both sets are explained in detail below and summarized in Table \ref{tab:res_grid_par}.

For a given mass scheme in our mass exploration, we produce {\N} simulations for each resonant configuration, initialize the planets on coplanar, circular orbits, 1\% wide of their assigned commensurability, and draw each planet's initial longitude from a uniform distribution between $0$ and $2\pi$. We produce four simulation suites of this form, each corresponding to a different planet mass scheme (labeled $\mathcal{S}_1$, $\mathcal{S}_2$, $\mathcal{S}_3$, and $\mathcal{S}_4$ in Table \ref{tab:res_grid_par}). The mass schemes include two equal mass and two hierarchical configurations; these schemes are motivated by \cite{Weiss2018}, which demonstrated general uniformity in a system's planet sizes and identified outer planets as statistically larger in systems of three or more planets. For each model, we fix the stellar mass (0.93 $M_{\odot}$) and initial period of the inner planet (7.1320 days), as well as $t_1$ and $t_2$, the lengths of the migration and tidal-damping phases, respectively. To ensure all systems capture into their assigned MMR, we set $\tau_a=10^8$~days and $K=10^3$, where $K \equiv \tau_a / \tau_e$; we explore the effects of changing the migration timescales in \S \ref{sec:tau_dep}. These parameters are summarized in Table \ref{tab:res_grid_par}. 

In our timescale investigation, we produce models on a grid of $\tau_a$ and $K$ values for each resonant configuration; we assign twenty-two $\tau_a$ values between $10^6$ and $10^9$ days and twenty-two $K$ values between $5$ and $10^3$, both in logarithmic spacing. Each planet's orbital elements are chosen as above, except that initial orbital periods are drawn from a uniform distribution between $0.5\%$ and $5\%$ wide of the assigned commensurability. The stellar mass, innermost orbital period, and $t_1$, $t_2$ are unchanged from the mass exploration suites ($\mathcal{S}_1$ through $\mathcal{S}_4$). For the planet masses, we adopt the 4.0, 4.0, 4.0 $M_{\oplus}$ scheme from $\mathcal{S}_1$. These models are summarized in Table \ref{tab:res_grid_par} and designated $\mathcal{S}_5$.

\begin{deluxetable*}{cccrrrrrrr}[t]
\tablecaption{Three-body Angle Equilibria for $\mathcal{S}_1$\label{tab:equ_mass}}
\tablehead{
\colhead{$P_2/P_1$} & \colhead{$P_3/P_2$} & \colhead{Three-body Angle} & \colhead{Center ($^{\circ}$)} &
\colhead{Amplitude ($^{\circ}$)} & \colhead{Fraction}
}
\startdata
2:1 & 2:1 & 1$\lambda_1$ - 3$\lambda_2$ + 2$\lambda_3$ & $180.08\pm0.01$ & $0.03\pm0.03$ & 30/30\\
\hline
2:1 & 3:2 & 1$\lambda_1$ - 4$\lambda_2$ + 3$\lambda_3$ & $180.03\pm0.02$ & $0.14\pm0.11$ & 30/30\\
\hline
2:1 & 4:3 & 1$\lambda_1$ - 5$\lambda_2$ + 4$\lambda_3$ & $180.01\pm0.03$ & $0.09\pm0.26$ & 30/30\\
\hline
2:1 & 5:4 & 1$\lambda_1$ - 6$\lambda_2$ + 5$\lambda_3$ & $179.98\pm0.06$ & $1.36\pm0.04$ & 30/30\\
\hline
3:2 & 2:1 & 2$\lambda_1$ - 4$\lambda_2$ + 2$\lambda_3$ & $\textbf{------}$ & $\textbf{------}$ & 30/30\\
\hline
3:2 & 3:2 & 2$\lambda_1$ - 5$\lambda_2$ + 3$\lambda_3$ & $180.02\pm0.05$ & $0.07\pm0.05$ & 30/30\\
\hline
3:2 & 4:3 & 2$\lambda_1$ - 6$\lambda_2$ + 4$\lambda_3$ & $162.00\pm0.12$ & $0.27\pm0.07$ & 11/30\\
 & & & $197.90\pm0.10$ & $0.21\pm0.12$ & 19/30\\
\hline
3:2 & 5:4 & 2$\lambda_1$ - 7$\lambda_2$ + 5$\lambda_3$ & $180.00\pm0.02$ & $0.63\pm0.02$ & 30/30\\
\hline
4:3 & 2:1 & 3$\lambda_1$ - 5$\lambda_2$ + 2$\lambda_3$ & $\textbf{------}$ & $\textbf{------}$ &  30/30\\
\hline
4:3 & 3:2 & 3$\lambda_1$ - 6$\lambda_2$ + 3$\lambda_3$ & $148.73\pm0.01$ & $0.24\pm0.03$ & 7/30\\
 & & & $180.00\pm0.03$ & $0.28\pm0.02$ & 16/30\\
 & & & $211.29\pm0.02$ & $0.25\pm0.03$ & 7/30\\
\hline
4:3 & 4:3 & 3$\lambda_1$ - 7$\lambda_2$ + 4$\lambda_3$ & $180.00\pm0.01$ & $0.41\pm0.02$ & 30/30\\
\hline
4:3 & 5:4 & 3$\lambda_1$ - 8$\lambda_2$ + 5$\lambda_3$ & $180.00\pm0.03$ & $0.73\pm0.06$ & 30/30\\
\hline
5:4 & 2:1 & 4$\lambda_1$ - 6$\lambda_2$ + 2$\lambda_3$ & $\textbf{------}$ & $\textbf{------}$ & 30/30\\
\hline
5:4 & 3:2 & 4$\lambda_1$ - 7$\lambda_2$ + 3$\lambda_3$ & $180.02\pm0.06$ & $0.45\pm0.36$ & 30/30\\
\hline
5:4 & 4:3 & 4$\lambda_1$ - 8$\lambda_2$ + 4$\lambda_3$ & $166.41\pm0.02$ & $0.58\pm0.06$ & 6/30\\
 & & & $174.54\pm0.03$ & $0.79\pm0.02$ & 12/30\\
 & & & $185.42\pm0.03$ & $0.78\pm0.03$ & 7/30\\
 & & & $193.58\pm0.01$ & $0.64\pm0.05$ & 5/30\\
\hline
5:4 & 5:4 & 4$\lambda_1$ - 9$\lambda_2$ + 5$\lambda_3$ & $180.00\pm0.03$ & $1.00\pm0.06$ & 30/30\\
\hline
\enddata
% \tablecomments{Caption}
\end{deluxetable*}

\section{Simulation Results}
\label{sec:sim_res}

From our numerical integrations, we find $180^{\circ}$ is the preferred libration center for nearly all three-body angles. The notable exceptions are the (3:2, 4:3), (4:3, 3:2), and (5:4, 4:3) configurations, where the three-body angle admits equilibria off $180^{\circ}$; we discuss these three configurations in detail in \S \ref{sec:shifts}.

To illustrate this behavior, in Figure \ref{fig:tb} we present the evolution of the three-body angle equilibria for $\mathcal{S}_1$ (see Table~\ref{tab:res_grid_par} for model parameters). We also report the libration center and amplitude for each equilibria in Table \ref{tab:equ_mass}, again for $\mathcal{S}_1$. The reported values are taken from a 1~kyr window beginning when the inner period ratio has spread 0.5\% from the assigned commensurability. With the exception of the (5:4, 4:3) chain, we find the three-body angle does not evolve considerably with additional spreading. 

In $\mathcal{S}_1$, we find that for three resonant configurations --- (3:2, 2:1), (4:3, 2:1), and (5:4, 2:1)--- the three-body angle circulates. This is not universal for these chains and is a consequence of the selected $\tau_a$ and $\tau_e$ for $\mathcal{S}_1$ through $\mathcal{S}_4$. As discussed in \S \ref{sec:tau_dep}, when different migration timescales are adopted, all three of these chains admit librating solutions with $\phi=180^{\circ}$. Unless otherwise stated, librating solutions refer to systems in which the three-body angle librates; as stated in \S \ref{sec:intro}, libration of the three-body angle does not guarantee libration of the two-body angles.

For the resonant equilibria reported in Figure \ref{fig:tb} and Table \ref{tab:equ_mass}, the three-body angles are found by the formulation of \S \ref{sec:formulation} and the libration amplitudes are given by \citep{Millholland2018},
\begin{equation}
\label{equ:ampl}
A=\sqrt{\frac{2}{N}\sum_{i=1}^N(\phi_i-\bar{\phi})^2}
\end{equation}
which approximates the amplitude of sinusoidal oscillations over a sample of $\phi$; $i$ indexes over the time-steps, $N$ is the number of time-steps, and $\bar{\phi}$ is the mean of the sample.

We next investigate the sensitivity of these equilibria to different mass schemes (\S \ref{sec:mass_dep}) and migration timescales (\S \ref{sec:tau_dep}), as well as discuss the off $180^{\circ}$ equilibria (\S \ref{sec:shifts}) and cases of three-body angle circulation (\S \ref{sec:nolib}).

\begin{deluxetable*}{ccrrrrrr}
\tablecaption{Three-body Angle Mass Dependency\label{tab:compare}}
\tablehead{
\colhead{$P_2/P_1$} & \colhead{$P_3/P_2$} & \colhead{$\bar{\phi}^{(\mathcal{S}_1)} - \bar{\phi}^{(\mathcal{S}_2)}$} &
\colhead{A$^{(\mathcal{S}_1)}$ $-$ A$^{(\mathcal{S}_2)}$} & \colhead{$\bar{\phi}^{(\mathcal{S}_1)} - \bar{\phi}^{(\mathcal{S}_3)}$} &
\colhead{A$^{(\mathcal{S}_1)}$ $-$ A$^{(\mathcal{S}_3)}$}
& \colhead{$\bar{\phi}^{(\mathcal{S}_1)} - \bar{\phi}^{(\mathcal{S}_4)}$} &
\colhead{A$^{(\mathcal{S}_1)}$ $-$ A$^{(\mathcal{S}_4)}$}
}
\startdata
3:2 & 4:3 & $0.33\pm0.24$ & $-0.13\pm0.22$ & $-0.18\pm0.12$ & $-0.09\pm0.07$ & $0.03\pm0.12$ & $-0.22\pm0.08$\\
 & & $-0.38\pm0.27$ & $-0.34\pm0.18$ & $0.11\pm0.11$ & $-0.16\pm0.12$ & $-0.08\pm0.11$ & $-0.27\pm0.12$\\
4:3 & 3:2 & $0.71\pm0.03$ & $-0.24\pm0.05$ & $0.13\pm0.01$ & $0.12\pm0.12$ & $0.32\pm0.01$ & $-0.07\pm0.03$\\
 & & $-0.01\pm0.03$ & $-0.24\pm0.02$ & $-0.01\pm0.03$ & $0.05\pm0.02$ & $-0.01\pm0.03$ & $-0.08\pm0.02$\\
 & & $-0.71\pm0.02$ & $-0.25\pm0.04$ & $-0.11\pm0.02$ & $0.06\pm0.03$ & $-0.34\pm0.02$ & $-0.06\pm0.03$\\
5:4 & 4:3 & $-0.39\pm0.08$ & $-0.59\pm0.15$ & $-2.98\pm4.29$ & $0.01\pm0.13$ & $-0.08\pm0.07$ & $-0.18\pm0.09$\\
 & & $-0.04\pm0.06$ & $-0.70\pm0.05$ & $-0.06\pm0.04$ & $0.08\pm0.04$ & $0.01\pm0.25$ & $-0.22\pm0.07$\\
 & & $0.33\pm0.29$ & $-0.63\pm0.10$ & $0.01\pm0.04$ & $0.07\pm0.03$ & $0.05\pm0.05$ & $-0.27\pm0.03$\\
 & & $-0.98\pm3.33$ & $-0.53\pm0.13$ & $-1.44\pm3.33$ & $0.21\pm0.18$ & $-1.30\pm3.33$ & $-0.11\pm0.08$\\
\enddata
\tablecomments{The three-body angle equilibria are presented in the same order as Table \ref{tab:equ_mass}, and suites $\mathcal{S}_1$ through $\mathcal{S}_4$ are described in Table~\ref{tab:res_grid_par}.}
\end{deluxetable*}

\begin{deluxetable}{cccc}
\tablecaption{Three-body Angle Equilibria for $\mathcal{S}_5$ (see Table \ref{tab:res_grid_par} for comparison)\label{tab:validation}}
\tablehead{
\colhead{Resonant Configuration} &
\colhead{Libration Center ($^{\circ}$)}
}
\startdata
(3:2, 4:3) & $162.08\pm0.60$\\
 & $197.88\pm0.49$\\
 \hline
(4:3, 3:2) & $148.80\pm0.11$\\
 & $180.01\pm0.39$\\
 & $211.41\pm0.27$\\
 \hline
(5:4, 4:3) & $167.24\pm0.78$\\
 & $173.56\pm2.30$\\
 & $185.52\pm1.88$\\
 & $192.56\pm0.67$\\
 \hline
Others & $180.09\pm0.42$\\
\enddata
\end{deluxetable}

\subsection{Mass Dependency}
\label{sec:mass_dep}
By comparing the three-body angle equilibria of $\mathcal{S}_1$ through $\mathcal{S}_4$, we investigate the sensitivity of the equilibria to planet masses. In general, the three-body angles of $\mathcal{S}_1$, $\mathcal{S}_2$, $\mathcal{S}_3$, and $\mathcal{S}_4$ are highly similar. We again find that 180$^{\circ}$ is the preferred libration center and identify shifts about $180^{\circ}$ in the same three resonant configurations. In Table \ref{tab:compare}, we compare the libration centers and libration amplitudes for the chains not centered at 180$^{\circ}$ between the four mass schemes. For ease of comparison, the reported values are again taken from a 1~kyr window beginning when the inner period ratio has spread 0.5\% from the assigned commensurability. The effects of varying the mass scheme are minimal; the differences in amplitude are all less than 1$^{\circ}$ and the differences in libration center are less than 5$^{\circ}$. For the super-Earth to sub-Neptune regime, shifts in the three-body angle due to planet masses are thus considerably less significant than those induced by non-adjacent resonances.

We note two significant differences across the suites. Firstly, in the two hierarchical configurations ($\mathcal{S}_3$ and $\mathcal{S}_4$), the three-body angle for the (3:2, 2:1) chain librates in all {\N} models about $\phi=180^{\circ}$, while it does not librate in any of the models for the two equal-mass suites ($\mathcal{S}_1$ and $\mathcal{S}_2$). Secondly, in the (5:4, 3:2) and (5:4, 5:4) chains, we find that only a fraction of the models librate in the three additional mass schemes ($\mathcal{S}_2$, $\mathcal{S}_3$, and $\mathcal{S}_4$), even though all the models librate in the 4.0, 4.0, 4.0 $M_{\oplus}$ scheme. 

As such, we find the equilibria positions are weakly dependant on the mass scheme, but whether the system librates or circulates can strongly depend on the mass scheme. 

\subsection{Migration Timescale Dependency}
\label{sec:tau_dep}
Using $\mathcal{S}_5$, we next explore the three-body angle equilibria for an array of migration timescales and initial positions. For each resonant configuration, we integrate synthetic systems for a grid of $\tau_a$ and $K$ values, with the initial orbital periods drawn from a uniform distribution between 0.5\% and 5\% wide of the assigned commensurability.

In Table \ref{tab:validation}, we report the libration centers for all systems that capture into their assigned commensurability and achieve steady-state by the end of the migration phase. As in Tables \ref{tab:equ_mass} and \ref{tab:compare}, the reported values are taken from a 1~kyr window beginning when the inner period ratio has spread $0.5\%$ from the assigned commensurability.

We find agreement between the libration centers reported in Table \ref{tab:validation} and those presented in Table \ref{tab:equ_mass}, indicating a weak dependence between the three-body angle equilibria and the migration timescales. However, we do not anticipate an exact match, since the range of $\tau_e$ values for $\mathcal{S}_5$ allows for larger libration amplitudes, and by extension, greater scatter in the reported three-body angles. 

\begin{figure*}[t]
\gridline{
\fig{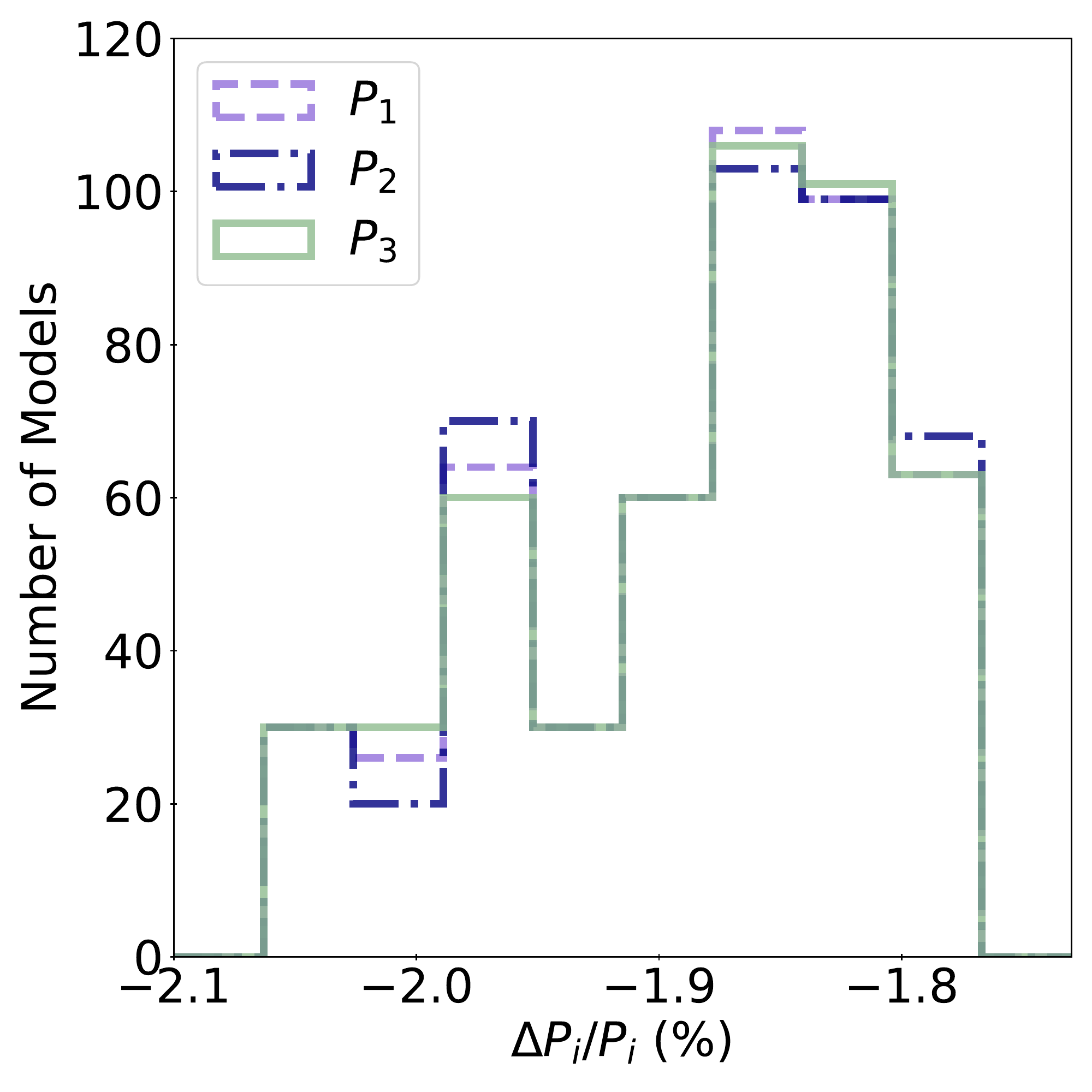}{0.333\textwidth}{}
\fig{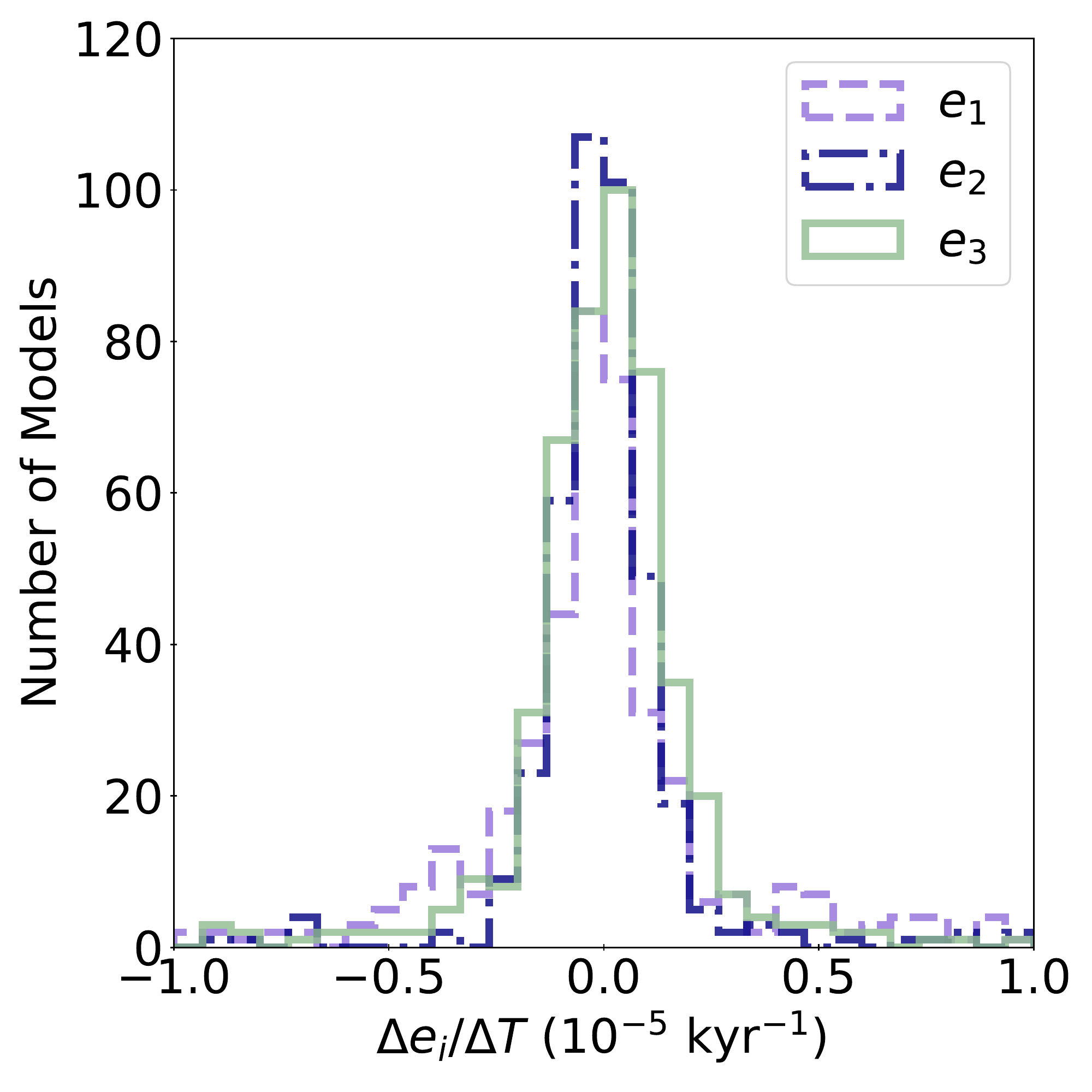}{0.333\textwidth}{} \fig{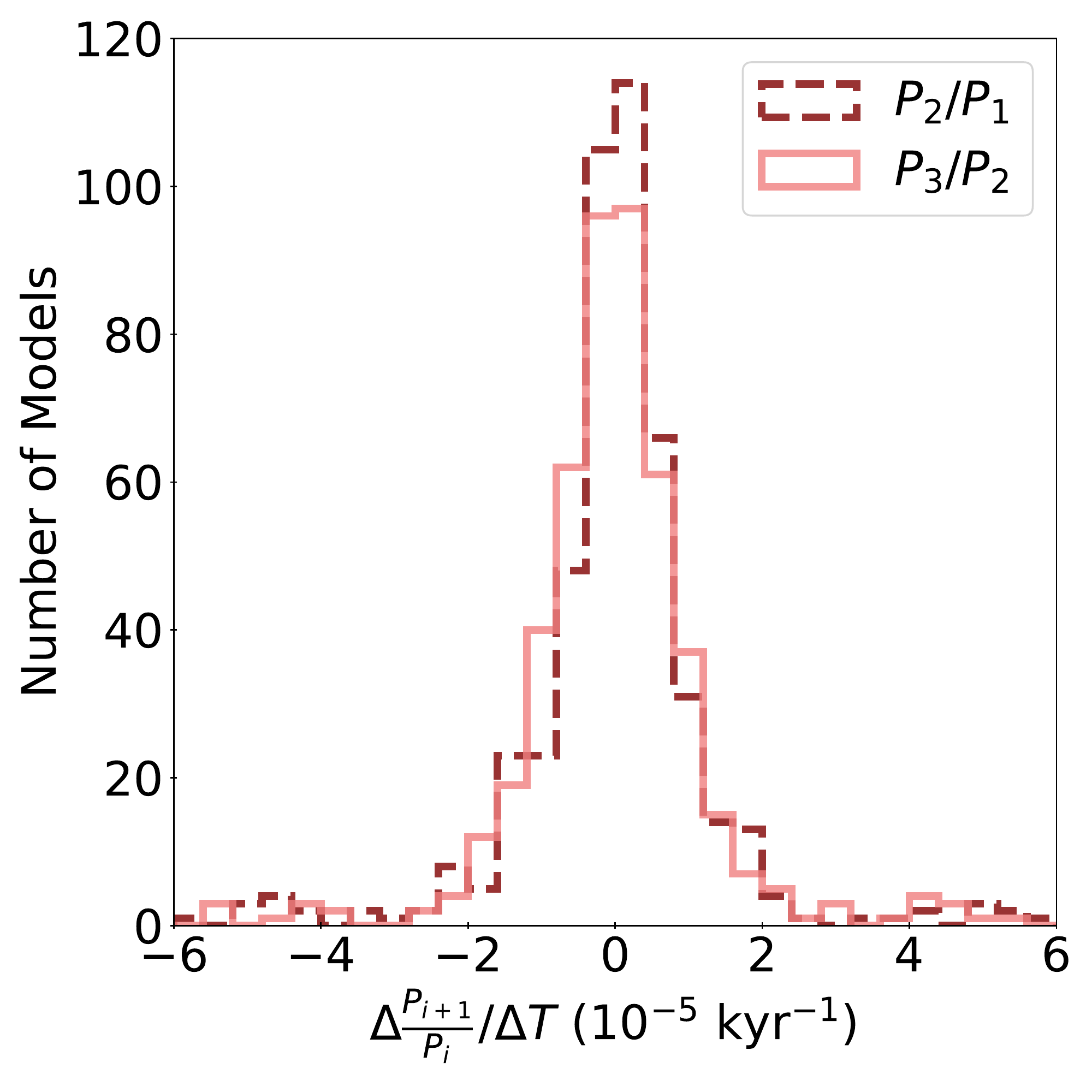}{0.333\textwidth}{}
}
\caption{Demonstration that all systems in $\mathcal{S}_1$ through $\mathcal{S}_4$, including those not librating, reach a steady-state during the last $10$~kyr of migration. \textit{Left}: the percent change in orbital period for each planet over the last 10~kyr of migration. \textit{Middle}: the rate of change in eccentricity for each planet over the same 10~kyr. \textit{Right}: the rate of change in the period ratios, again over the last 10~kyr of migration.}
\label{fig:end_of_migr}
\end{figure*}

\subsection{Off $180^{\circ}$ Equilibria}
\label{sec:shifts}

Across our models, we find 180$^{\circ}$ is the preferred equilibria for the three-body angle, with the exception of three resonant configurations: (3:2, 4:3), (4:3, 3:2), and (5:4, 4:3).

In the (3:2, 4:3) and (4:3, 3:2) chains, the deviations from 180$^\circ$ occur at the moment libration begins and show little evolution in the tidal-dissipation regime. From Tables \ref{tab:equ_mass} and \ref{tab:validation}, we find the shifts are symmetric about 180$^\circ$ for a wide range of migration timescales and initial conditions. This behavior was predicted by \citet{delisle2017}, because for both configurations there is a first-order non-adjacent MMR, $P_3:P_2\approx2:1$.

Unlike the (3:2, 4:3) and (4:3, 3:2) chains, shifts in the three-body angle for the (5:4, 4:3) configuration do not fully present themselves until the tidal-dissipation regime; however, these libration centers still obey the near-symmetry about $180^{\circ}$. Since  $P_3:P_2\approx5:3$, this behavior was not considered by the semi-analytic model of \citet{delisle2017}, which was limited to first-order non-adjacent MMR.

Prompted by the (5:4, 4:3) chain, where the three-body angles continue spreading in the tidal-dissipation regime, we repeat $\mathcal{S}_1$ for the (3:2, 4:3), (4:3, 3:2), and (5:4, 4:3) chains using a range of $Q^{\prime}$ values. Consistent with \cite{Papaloizou2015}, we find that lowering the assumed $Q^{\prime}$ value by an order of magnitude corresponds to an order of magnitude speedup in the evolution. In the lower $Q'$ suites, the three-body angles of the (3:2, 4:3), (4:3, 3:2), and (5:4, 4:3) chains all clearly spread in the tidal-dissipation regime, while only the (5:4, 4:3) configuration visibly spreads in Figure \ref{fig:tb}. The effect is strongest in the (5:4, 4:3) chain and significantly weaker in the (3:2, 4:3) and (4:3, 3:2) chains. 

Thus, we are able to expand upon the semi-analytic findings of \citet{delisle2017} and conclude that not only are higher-order non-adjacent resonances capable of off $180^{\circ}$ equilibria, but their dependence on tidal-damping appears significantly amplified. As a result, for systems near a (5:4, 4:3) chain, such as Kepler-60, the three-body angle can occupy a range of values, depending on the evolutionary history of the system, see \S \ref{sec:k60} for discussion. 

\subsection{Non-librating Three-body Angles}

\label{sec:nolib}

Although a full investigation into the systems with non-librating three-body angles is beyond the scope of this paper, here we briefly discuss the parameters that influence the libration state of the three-body angle and compare the migration histories of the librating and non-librating populations.

In \S \ref{sec:mass_dep}, we found the libration state of the three-body angle depends strongly on the mass scheme. Additionally, we found that for each resonant configuration, certain regions of the $\tau_a$ and $K$ space admit librating solutions, while others admit circulating ones (\S \ref{sec:tau_dep}). Given the stochastic nature of this parameter space, the large number of resonant configurations, and the strong mass dependence, we refrain from characterizing these regions of libration and circulation and leave this analysis to future work.

We also find that for the systems with non-librating three-body angles in $\mathcal{S}_1$ through $\mathcal{S}_4$, each planet still captures into its assigned commensurability and reaches eccentricity-equilibrium by $t_1$. This is evidenced by Figure \ref{fig:end_of_migr}, where we present the change in the periods, period ratios, and eccentricities over the final 10~kyr of the migration phase for all models in $\mathcal{S}_1$ through $\mathcal{S}_4$; the eccentricity and period ratio distributions are both consistent with zero, while the periods themselves are shrinking. This confirms the planets are migrating in lock-step, even for the systems with circulating three-body angles.

\section{Application}
\label{sec:application}

In transit data, three-body angles are often nontrivial to extract, complicating the application of our results to observed systems. Prior studies have often turned to forward integration of transit timing variation (TTV) solutions \citep{godz2016,MacDonald2016,Jontof-Hutter2016}, or considered the zeroth order in $e$ approximation of \citet{Mills2016} and \cite{2017Luger}. 

Here, we first derive this approximation and then use this method, along with forward integration of TTV solutions, to compare the three-body angles of Kepler-60 and Kepler-223 with the results of \S \ref{sec:sim_res}.  For Kepler-60, we find the TTV solutions agree with our results and suggest the system has undergone minimal tidal-dissipation, while for Kepler-223, we update the three-body angles reported in \citet{Mills2016}. For both systems, we discuss the quality of the zeroth order in $e$ approximation of $\phi$ as a diagnostic of the three-body angle.

\subsection{Three-body Angles from TTV Data}
\label{sec:ttv}

A transit time $T$ can be related to a mean anomaly $\lambda$, by noting that the value of the true anomaly $f$ at mid-transit obeys $f + \omega = \pi/2$; this means the line between the star and the planet has swept $\pi/2$ from crossing the sky plane to the time of mid-transit\footnote{The appendix of \cite{2019Hamann} showed that when the planet is eccentric and inclined from edge-on, an additional term $\Delta T / P \propto (i-\pi/2)^2 e \cos \omega$ is needed, which we neglect here.}. The relation between $f$ and time is, to second order in $e$ \citep[eq. 2.88, 2.93]{1999MurrayDermott}:
\begin{eqnarray}
    f &=& M + 2 e \sin M + (5/4) e^2 \sin 2 M ,\\
    M &=& f - 2 e \sin f + (3/4) e^2 \sin 2 f , \label{eqn:Mf}
\end{eqnarray}
where $M= \lambda-\varpi$. Here we are setting $\Omega=0$, so $\varpi = \omega$. We may then take equation~\ref{eqn:Mf} and trigonometric identities to determine $\lambda$ at the mid-time of a transit: 
\begin{eqnarray}
    \lambda_t &=& \pi/2 -2e \cos \omega + 3/4 e^2 \sin 2 \omega, \\
    \lambda_t &=& \pi/2 - 2(e\cos \omega) + 3/2 (e \sin \omega) (e \cos \omega),
\end{eqnarray}
again to second-order in eccentricity. Since $\lambda_t$ is a \emph{mean} longitude, which progresses linearly in time, we may express it as a function of transit time and period: 
\begin{equation}
    \lambda_t = \frac{\pi}{2} + 2 \pi \frac{t-T_n}{P} - 2e\cos \omega  + \frac32 (e \sin \omega) (e \cos \omega) 
\end{equation}
Combining for all three planets, we have: 
% \begin{eqnarray}
% \phi &=& p \lambda_1 - (p+q) \lambda_2 + q \lambda_3, \\
% \phi_{\rm ttv} &=& 2\pi \Big(p  \frac{t-T_1}{P_1} - (p+q) \frac{t-T_2}{P_2} + q \frac{t-T_3}{P_3} \Big) \label{eqn:phitrans} \\
% \phi-\phi_{\rm ttv}&=& p(-2e_1\cos \omega_1 + \frac32 e_1^2\sin \omega_1\cos \omega_1) \nonumber \\
% & & -(p+q)(-2e_2\cos \omega_2 + \frac32 e_2^2\sin \omega_2\cos \omega_2)  \nonumber \\
% & & + q(-2e_3\cos \omega_3 + \frac32 e_3^2\sin \omega_3\cos \omega_3) \nonumber \\
% & & + \mathcal{O}(e_1^3,e_2^3,e_3^3).\label{eqn:phicorrector}
% \end{eqnarray}
\begin{eqnarray}
% \phi &=& j \lambda_1 - (j+k+1) \lambda_2 + (k+1) \lambda_3, \\
\phi_{\rm ttv} &=& 2\pi \Big(j  \frac{t-T_1}{P_1} - (j+k+1) \frac{t-T_2}{P_2} \nonumber \\
& &\qquad +(k+1) \frac{t-T_3}{P_3} \Big) \label{eqn:phitrans} \\
\phi-\phi_{\rm ttv}&=& j(-2e_1\cos \omega_1 + \frac32 e_1^2\sin \omega_1\cos \omega_1) \nonumber \\
& & -(j+k+1)(-2e_2\cos \omega_2 + \frac32 e_2^2\sin \omega_2\cos \omega_2)  \nonumber \\
& & + (k+1)(-2e_3\cos \omega_3 + \frac32 e_3^2\sin \omega_3\cos \omega_3) \nonumber \\
& & + \mathcal{O}(e_1^3,e_2^3,e_3^3).\label{eqn:phicorrector}
\end{eqnarray}
where $\phi_{\rm ttv}$ is the three-body angle inferred using $\lambda_t$ to zeroth order in $e$, $\phi$ is defined in equation \ref{equ:tb}, and $j,k$ are defined as in \S 1. Previous publications (e.g., \citealt{2017Luger}) took $e\rightarrow 0$ and only considered $\phi_{\rm ttv}$, since there is no easy observational access to $e$ and $\omega$, like there is for $T$ and $P$. However, since we are interested in comparing the three-body angles of transiting systems with our synthetic systems, we also consider the higher order correction $\phi-\phi_{\rm ttv}$.

\begin{figure*}[]
\centering
\includegraphics[width=.9\textwidth]{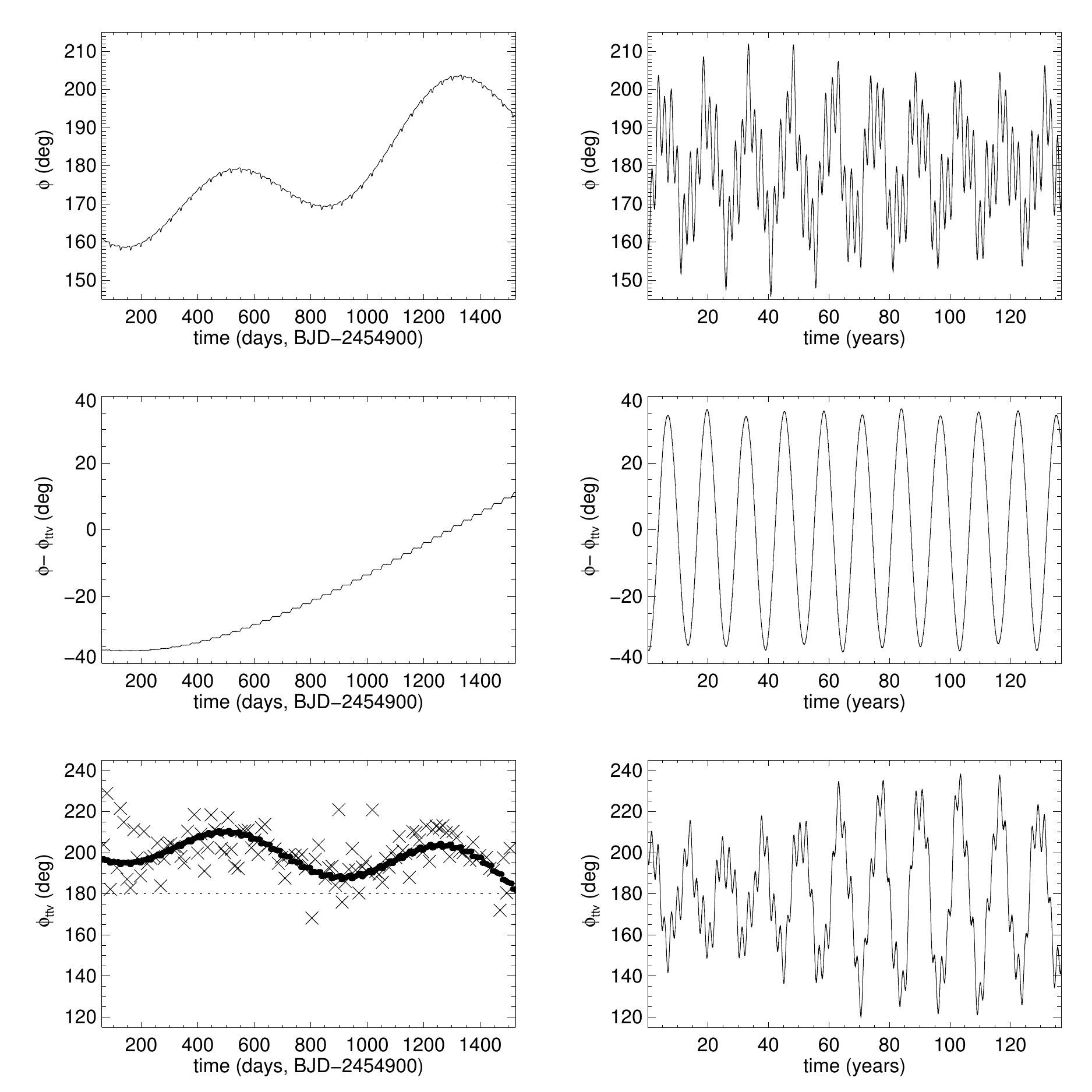}
\caption{The evolution of Kepler-60's three-body angle over time, from a fit to the data \citep{Jontof_Hutter_2021}. In the top two panels, the exact three-body angles are displayed, over the course of the data and over the next 137 years. In the middle panels, the difference between the true angles and the transit-identifiable angles is shown. In the bottom panels, the transit-identifiable angles (i.e., \citealt{2017Luger}, and eqn.~\ref{eqn:phitrans} of this paper) are shown. The values from the transit-timing data are overplotted in the bottom-left panel, measured at planet d's transit times using the closest transits of planets b and c. An apparent libration around $200^\circ$ is illusory, due to short-timescale fluctuations; the long-term behavior in the bottom-right panel shows the expected libration near $180^\circ$ (actually $179.12^\circ$, due to the 5:3 resonance between planets b and d). }
\label{fig:ttv}
\end{figure*}

\subsection{Kepler-60}
\label{sec:k60}
Kepler-60 hosts three transiting super-Earths with periods near a chain of first-order commensurabilities, 5:4 for the inner pair and 4:3 for the outer pair \citep{Steffen2013}. Through forward integration of TTV solutions, \cite{godz2016} suggested the system's three-body angle was librating and the two-body angles were likely librating as well. According to a recent erratum, \cite{Jontof_Hutter_2021} also found the TTVs implied a librating three-body angle. Both studies quote a libration center of $\sim45^\circ$; our standard three-body angle definition multiplies their definition by $4$, corresponding to $\phi\approx180^{\circ}$ with an amplitude of $~20^{\circ}$. 

Using three dynamical histories ---long-scale migration, short-scale migration, and pure eccentricity damping--- \cite{MacDonald2018} simulated the formation of Kepler-60 and found four equilibria for the three-body angle, none of which are consistent with $\phi=180^{\circ}$; in our standard formulation, these equilibria are approximately  $164.52-178.72 ^{\circ}$, $176.52-179.36 ^{\circ}$, $180.12-185.64 ^{\circ}$, and $182.36-189.92 ^{\circ}$. We find general agreement between our results for the (5:4, 4:3) configuration and the modeling by \cite{MacDonald2018}, but given the significant dependence of the libration centers on tidal-dissipation in the (5:4, 4:3) chains (see \S \ref{sec:shifts}) and the range of damping timescales employed by \cite{MacDonald2018}, we do not expect an exact match.

We revisit the 50 samples described in \citet[section 3.5]{Jontof-Hutter2016}, available via \cite{Jontof_Hutter_2021}, integrating them 1000 years from the start of the Kepler data ($t=0$ is BJD 2454900), to infer $\phi$.  We analyze an interval between $t=100$~yr and $1000$~yr, so that the phase of $\phi$ loses coherence among the various samples from their starting condition fitting the data. We find an average value of $\phi=179.12^\circ\pm0.10^\circ$, with a well-constrained amplitude of $17.5^\circ\pm2.9^\circ$.

As shown in Figure \ref{fig:tb}, our synthetic systems predict the three-body angle of the (5:4, 4:3) configuration spreads from $180^{\circ}$ through tidal-dissipation. We verified that the value of $179.12^\circ$ measured from the data posterior above is typical of the centers observed in $S_5$ when the period ratios match that of the observed system. 

From the match between theory and observation for the libration center of $\phi$, we conclude Kepler-60 has undergone limited tidal-damping. This agrees with \cite{Papaloizou2015}, which estimated $Q^{\prime}\gtrapprox 67,000$.

Using TTV data from \cite{2015Rowe}, we present the zeroth order in $e$ three-body angle of Kepler-60 in Figure \ref{fig:ttv}. We find this zeroth order approximation of $\phi$ seems to librate around $200^{\circ}$, with an amplitude of $10^{\circ}$. Thus, over the \textit{Kepler} observing window, the libration center of the zeroth order $\phi$ appears discrepant with our models (see Table \ref{tab:equ_mass}) and the TTV forward integrations of \citet{godz2016} and \citet{Jontof_Hutter_2021}. However, since the system is only slightly wide of near-integer ratios, we expect the higher-order corrections in $e$ to be non-negligible. For Kepler-60, the first-order in $e$ correction is: 
\begin{equation}
-8e_b\cos \omega_b +16 e_c \cos \omega_c -8 e_d \cos \omega_d
\end{equation}
and the second-order error term is: 
\begin{eqnarray}
& &6 (e_b\cos \omega_b)(e_b\sin \omega_b)  \nonumber \\
& &-12 (e_c \cos \omega_c)(e_c \sin \omega_c) \nonumber \\
& &+6 (e_d \cos \omega_d)(e_d \sin \omega_d).
\end{eqnarray}

In Figure \ref{fig:ttv}, we present the exact three-body angle ($\phi$) and the difference between $\phi$ and the zeroth order approximation ($\phi-\phi_{ttv}$) for a single draw from the posterior \citep{Jontof-Hutter2016,Jontof_Hutter_2021}. The apparent libration near $200^\circ$ just reflects short-period components of the total oscillation. Meanwhile, the corrector term librates with a period of $\sim30$~yr and an amplitude of $30^{\circ}$. From this forward integration, it appears the \textit{Kepler} observing window was coincident with a near maximal discrepancy between the exact three-body angle and the zeroth order approximation.

Although both $\phi$ and $\phi_{\rm ttv}$ agree with our models on longer timescales ($>15$~yr), for an arbitrary 1500~day observing window this is not guaranteed. On such short timescales, the discrepancy between $\bar{\phi}$ and $\bar{\phi}_{\rm ttv}$ may be non-negligible and neither angle may reflect their true libration centers. We thus urge caution when comparing the zeroth order approximation from TTV data to either our results or the semi-analytical solutions of \citet{delisle2017}.

\subsection{Kepler-223}

Kepler-223 hosts four transiting planets with periods near a (4:3, 3:2, 4:3) resonant chain \citep{Rowe2014}. Using the zeroth order approximation outlined in \S \ref{sec:ttv}, \cite{Mills2016} reported libration in $\phi_{1,ttv,M16}$ between 173$^{\circ}$ and 190$^{\circ}$ and in $\phi_{2,ttv,M16}$ between 47$^{\circ}$ and 75$^{\circ}$; under our standard formulation, $\phi_{1,ttv,M16}$ is multiplied by $-3$ and $\phi_{2,ttv,M16}$ is multiplied by 2, corresponding to 150$^{\circ}-201^{\circ}$ and 94$^{\circ}-150^{\circ}$, respectively. 

Using the same formulations as \cite{Mills2016}, \cite{delisle2017} predicted six equilibria for Kepler-223, including the values reported by \cite{Mills2016}. Unlike our three-planet models, where at most one non-adjacent resonant pair was possible, the (4:3, 3:2, 4:3) resonance configuration of Kepler-223 introduces two pairs of first-order resonances between non-adjacent planets --- $P_d : P_b \approx 2:1$ and $P_e : P_c \approx 2:1$ --- as well as a higher-order resonance between non-adjacent planets --- $P_e : P_b \approx 8:3$. Under the \cite{delisle2017} model, these resonances result in numerous stable equilibria. The closest equilibria to the $\phi_{ttv}$ reported by \cite{Mills2016}, translated into our parameters, is $\phi_{1}=168^\circ$ and $\phi_{2} = 130^\circ$ \citep{delisle2017}.

Here, recognizing from equation~\ref{eqn:phicorrector} that there may be a significant offset between $\phi$ and $\phi_{ttv}$ due to eccentricity, we revisit the Kepler-223 analysis. In \cite{Mills2016} Extended Data Figure 5, it is shown that not all the systems that fit the data are stable, and not even all the ones that had been selected for 100~year $\phi$ libration (prior $\mathcal{C}_3$) maintain the property that is so characteristic of the system: that period ratios lie within $\sim 0.001$ of an integer ratio. In particular, from the quarter-to-quarter timing analysis, the period ratios of adjacent pairs fall in the ranges $P_c/P_b \in [1.3326,1.3341]$, $P_d/P_c \in [1.5007,1.5025]$, and $P_e/P_d \in [1.3331,1.3346]$. To enforce that we are studying the $\phi$ behavior of systems that stay comparably close to integer ratios, we inspect the 2,008 runs that show stability for at least 1~Myr, from a draw of 5,000 systems fitting the data (the $\mathcal{C}_2$ prior). We compute the fraction of time that all three period ratios are in the above ranges (we allow $P_d/P_c$ to go down to $1.500$), i.e. very close to commensurability. A bimodality exists, in which 99.3\% of the systems are this close to commensurability less than 18\% of the time, with period ratios typically straying $\gtrsim0.2$~\% from commensurability; 0.7~\% of the systems, just 14 of the 2,008, stay in these ranges $>29$\% of the time. 

Thus we would expect that if the period ratios of Kepler-223 really do wander, and are found so close to commensurability during the Kepler observations by chance, then $\gtrsim5$ ``clones'' of the system would be found with period ratios within $\sim0.5\%$ of commensurability, but not within the bounds above. Recently, \cite{2020Jiang} curated a list of near resonant systems from \textit{Kepler}, however, none of them are consistent with this characterization. This population consideration justifies the selection of only the systems that maintain period ratios very close to integer ratios. 

With our selected data-fitting and stable systems, that maintain small period ratio variations, we find the libration center and amplitude (equation~\ref{equ:ampl}) of each $\phi$, over $10^6$ years: 
$\phi_1=165^{+11}_{-5}$~$^{\circ}$, $A_1=17^{+8}_{-3}$~$^{\circ}$ and $\phi_2=133^{+15}_{-7}$~$^{\circ}$, $A_2=30^{+68}_{-10}$~$^{\circ}$. These central values are the median of the 14 values, and the error bars are the distance to the second and second-to-last values of the ordered set of 14 values, approximately 1-$\sigma$. These results depend on the dynamical model of the system, so it should not be surprising that they substantially agree with \cite{delisle2017}, based on the values quoted above. 

The motivation for defining $\phi_{\rm ttv}$ is that no dynamical model needs to be invoked, so it is a nearly direct observable. However, there are higher order in $e$ corrections that should be estimated from the dynamical models (equation~\ref{eqn:phicorrector}). For the set of 14 selected systems, the correction for $\phi_1$ scans from $\sim -40^\circ$ at the start of the Kepler dataset to $\sim 5^\circ$ at the end and from $\sim 10^\circ$ to $-25^\circ$ for $\phi_2$. These values are larger than the errors quoted above, therefore we point out the importance of consulting dynamical models when determining $\phi$, rather than relying solely on $\phi_{\rm ttv}$, despite its utility in being observable and easily computed. 

\section{Conclusions}
\label{sec:concl}

Using tidally-damped N-body integrations, we characterized the three-body angle equilibria in compact three-planet systems for each combination of the 2:1, 3:2, 4:3, and 5:4 mean motion resonances (MMRs). We conducted this survey over an array of planet mass schemes, migration timescales, and initial separations. We then related these results to two transiting systems ---Kepler-60 and Kepler-223--- through transit timing variation (TTV) analysis. 

We found that under our formulation of the three-body angle, which does not reduce the coefficients of equation \ref{equ:tb}, $180^{\circ}$ is the preferred libration center and low-order, non-adjacent MMRs result in shifts about $180^{\circ}$, specifically in the (3:2, 4:3), (4:3, 3:2), and (5:4, 4:3) chains. In the chains with off $180^{\circ}$ libration centers, we also identified a relationship between the three-body angle and tidal-dissipation effects; this relation is strongest in the (5:4, 4:3) chain and is thus not characterized under the semi-analytic model of \cite{delisle2017}, which was restricted to first-order non-adjacent resonances. An exploration of this behavior is beyond the scope of this work, however, we encourage future study of this relationship. 

Through TTV analysis of Kepler-60 and Kepler-223, we considered the application of our results to transiting exoplanet systems. Comparison of the forward integrated TTV solutions with our results revealed that Kepler-60 has likely undergone minimal tidal-dissipation, consistent with the $Q^{\prime}$ estimation of \citealt{Papaloizou2015}. For both systems, we found that on short timescales the zeroth order in $e$ approximation for the three-body angle \citep[see][]{Mills2016,2017Luger} may be significantly discrepant from the three-body angles inferred via forward integration. As such, we stress the importance of consulting dynamical models when investigating three-body angles and caution against comparison of $\phi_{\rm ttv}$, the zeroth order approximation, to either our results or the models of \cite{delisle2017}.

We propose that our results provide a simple diagnostic of observed three-body angles and should guide future use of $\phi_{ttv}$. 

% [FUTURE WORK?]
%%%%%%%%%% CLOSING %%%%%%%%%%
\ \\
\acknowledgments
We thank the referee for their
comments, which have helped to substantially improve
the paper. We thank Daniel Jontof-Hutter, for helping us examine the Kepler-60 system, and Jack Lissauer, for discussing the long-term periods of Kepler-223. This work was completed in part with resources provided by the University of Chicago Research Computing Center. DF thanks former research assistants Kathryn Chapman, Sean Mills, and Enid Cruz-Col\'{o}n for their work on resonant chains. JS thanks the Hoeft College Research Fellowship at the University of Chicago, for financial support.

% \facilities{XMM(EPIC)}

\bibliography{paper}%

\end{document}